\documentclass[10pt]{article}
\usepackage{amsmath,amsfonts,amscd}
\usepackage{amsthm}
\newcommand{\as}{\mathfrak{P}}

\newcommand{\copos}{\mathbf{Sh}(X)}
\newcommand{\dd}{\text{\texttt{d}}}

\newcommand{\kodd}{\delta}

\newcommand{\qopos}{{}^{fcq}\copos}

\newcommand{\rz}{\mathfrak{R}}
\newcommand{\hil}{\mathfrak{H}}

\newcommand{\cont}{\mathcal{C}^{0}}
\newcommand{\smooth}{\mathcal{C}^{\infty}}
\newcommand{\latt}{\mathcal{U}\mathcal{P}}

\newcommand{\com}{\mathbb{C}}

\title{\bf Presheaves, Sheaves and their Topoi in\\ Quantum
Gravity and Quantum Logic\thanks{Short paper version of the talk
``{\itshape Reflections on a Possible `Quantum Topos' Structure
Where Curved Quantum Causality Meets `Warped' Quantum Logic}''
given at the {\sl 5th International Quantum Structures Association
Conference} in Cesena, Italy (31/3---5/4/2001).}}

\author{Ioannis Raptis\thanks{\rm EU Marie Curie Research Fellow,
Theoretical Physics Group, Blackett Physics Laboratory, Imperial
College of Science, Technology and Medicine, Prince Consort Road,
South Kensington, London SW7 2BZ, UK; i.raptis@ic.ac.uk}}

\date{}

\begin{document}
\maketitle

\begin{abstract}

\noindent A brief synopsis of recent conceptions and results, the
current status and future outlook of our research program of
applying sheaf and topos-theoretic ideas to quantum gravity and
quantum logic is presented.

\end{abstract}

\vskip 0.2in

\centerline{``{\em Logics come from dynamics}"}

\centerline{(D. R. Finkelstein: `{\sl Quantum Relativity}'
\cite{df96})}

\vskip 0.2in

\section{Introduction: two questions motivating our quest}\label{sec1}

The following two questions, one physical the other more
mathematical, motivate essentially our general research project of
applying sheaf and topos-theoretic concepts, techniques and
results to quantum gravity and quantum logic:

\begin{itemize}

\item {\em Is there a fundamental
connection between the quantum logical structure of the world and
its dynamically variable microcausal or chronological structure at
Planck scales as the latter is supposed to be determined by the
until now persistently elusive quantum theory of gravity?}

\end{itemize}

\noindent and related to it:

\begin{itemize}

\item {\em How can one localize noncommutatively?}

\end{itemize}

Concerning the first question, we intuit that a sound theoretical
scheme for quantum gravity should be intimately related to the
logical structure of the world at quantum scales: in a strong
sense, quantum causality should be unified at the dynamical level
with quantum logic in the light of quantum gravity. In turn, this
conjecture essentially implies our main suspicion that in the
quantum spacetime deep even quantum logic should be regarded as a
quantum `observable' entity that is subject to dynamical
changes---a {\em dynamical physical logic} analogous to the
dynamical physical spacetime geometry of the classical theory of
gravity ({\itshape ie}, general relativity) \cite{df69b, df79}.
That {\em logos is somehow related to chronos at a fundamental
level}\footnote{In fact, the opening quotation from \cite{df96}
suggests that {\em logic derives from dynamics}!} has become the
central theme in our quantum gravity research program over the
last few years.

Our subsequent decision to implement mathematically this theme by
using sheaf and topos-theoretic concepts, techniques and results
is based on the by now widely established fact, at least among
categorists and related `toposophers', that {\em the theory of
presheaves, sheaves and their topoi fuses geometry with logic at a
basic level} \cite{lawv75, gold77, macmo92}. It only appeared
natural to us that if geometry could be somehow identified with
`spacetime geometry' in particular, while logic with `quantum
logic', then the long sought after unification of relativity with
quantum mechanics could be possibly achieved by sheaf and
topos-theoretic means. After all, the methods of sheaf and topos
theory are of an essentially algebraic nature \cite{macmo92,
mall1, mall2}, and lately there has been a strong tendency among
mathematical physicists to tackle the problem of quantum gravity
entirely by categorico-algebraic means \cite{crane, mall3, mall4,
mallrapt00, mallrapt01}.

Concerning the second motivating question above which, as we will
argue subsequently, is closely related to the first, our quest
focuses on a possible formulation of a {\em noncommutative
topology} and its associated {\em noncommutative sheaf theory}
that can can be applied to the problem of the quantum structure
and dynamics of spacetime. Our original motivation for looking
into the possibility of a noncommutative or, ultimately, `quantum'
topology for quantum gravity rested heavily on our desire to
abandon the geometric spacetime continuum on which the mathematics
of general relativity ({\it ie}, the standard differential
geometry) essentially rests for some structure of a more
finitistic, algebraic and, hopefully, dynamical character
\cite{rapzap1, mallrapt00, rapzap2, mallrapt01, rap4}. This is the
subject of the next section and we will use it as the {\it raison
d'\^etre} of our endeavor to apply sheaf and topos-theoretic ideas
to quantum gravity.

Thus, the short report below commences with various physical and
mathematical evidence that we have collected in the past couple of
years against the classical topological ({\it ie},
$\cont$-continuous) and differential ({\it ie}, $\smooth$-smooth)
manifold model of spacetime by essentially basing ourselves on
quantum theory's principles of finiteness or discreteness,
superposition and, as a result, algebraic noncommutativity, as
well as on relativity's central principle of local causality
commonly known as `locality'. Thus, we will see how effectively
sheaf and topos-theoretic ideas may be used to formulate a {\em
locally finite, causal and quantal version of} (at least the
kinematics of) {\em discrete Lorentzian quantum gravity}
\cite{rapzap1, rap2, rap3, mallrapt00, rapzap2, mallrapt01, rap4,
rap6}, for it has been convincingly argued that capturing the
`proper' kinematical structure constitutes the first decisive step
towards arriving at the notoriously elusive quantum dynamics for
spacetime and gravity \cite{sork95, rapzap2}.

\section{The Past: manifold reasons against the spacetime
manifold}\label{sec2}

A noncommutative geometry \cite{connes94} has already been
proposed, significantly worked out and diversely applied to the
problem of the quantum structure and dynamics of spacetime ({\it
ie}, quantum gravity). However, it seems theoretically rather {\it
ad hoc}, lame and short sighted to think of a higher level
structure such as the geometry of spacetime as being subject to
some sort of quantization and as participating into, in principle
measurable, dynamical variations\footnote{That is, in general
relativity at least, the gravitational field, which is represented
by the spacetime metric $g_{\mu\nu}$, is treated as an {\em
observable}; in fact, the sole spacetime observable.} thus be
soundly modelled by noncommutative mathematics, while a more basic
structure such as the spacetime topology to be treated essentially
as a fixed classical entity, hence be modelled after a non-varying
locally Euclidean manifold equipped with algebras of commutative
coordinates labelling its point events \cite{rapzap1, mallrapt00,
rap4, rapzap2}. Related to this, and from a rather general and
technical perspective, while a commutative sheaf theory has been
rather quickly developed, well understood and widely applied to
both mathematics and physics \cite{harts, eisenhar, shaf94, mall1,
mall2, mall3, mall4}, a noncommutative one (and the topology
related to it) has been rather slow in coming and certainly not
unanimously agreed on how to be applied to quantum spacetime
research \cite{shaf94, voysta, voyve, mulpel1, mulpel2, bor2,
bor3, bor1, mallrapt00, rap4}\footnote{That is to say, not all
mathematicians and mathematical physicists agree on what ought to
qualify as `{\em noncommutative topology}' proper and its related
noncommutative sheaf or scheme theory. At the same time, there is
no collective agreement on how such a noncommutative or quantum
\cite{isham1, isham2, isham3, df912, df96} topology may be applied
to the problem of the quantum structure and dynamics of
spacetime.}.

At the same time, and from a physical point of view, the
unreasonableness and unphysicality of the locally Euclidean
topological ($\cont$) and differential ($\smooth$) manifold model
$M$ for spacetime is especially pronounced when one considers:

\begin{itemize}

\item (a) \underline{\bf Pointedness of events:} $M$'s pathological nature
in the guise of singularities that plague general relativity---the
classical theory of gravity---which are mainly due to the
geometric point-like character of the events that constitute it,
as well as due to the algebras of $\smooth$-smooth functions
employed to coordinatize these point events \cite{mallrapt01} (and
also due to (b) next).

\item (b) \underline{\bf Continuous infinity of events:} $M$'s
problematic nature due to the fact that one can in principle pack
an uncountable infinity of the aforementioned point events in a
finite spacetime volume resulting in the non-renormalizable
infinities that impede any serious attempt at uniting quantum
mechanics with general relativity (at least at the `calculational'
level)\footnote{The (a) and (b) pathological features of the
manifold model above may be summarized in its character as {\em a
geometric point set differential continuum of events}
\cite{mallrapt01}.}.

\item (c) \underline{\bf Non-dynamical and non-quantal topology:} Its
non-variable and non-quantal nature when one expects that at
Planck scales not only the spacetime metric, but also that the
spacetime topology partakes into quantum phenomena
\cite{wheeler64}, that is to say, it is a dynamically variable
entity whose connections engage into coherent quantum
superpositions. We may distill this by saying that the manifold
topology is, quantally speaking, an {\em unobservable} entity not
manifesting quantum dynamical fluctuations or interference between
its defining connections \cite{rapzap2, rap4}---a rigid substance,
once and forever fixed by the theorist, that is not part of the
dynamical flux of Nature at microscopic scales. Furthermore, the
(algebras of) commutative $\smooth$-determinations of the
manifold's point events indicate another non-quantal (classical)
feature of the spacetime manifold \cite{rapzap1, mallrapt01}.

\item (d) \underline{\bf Additional structures:} $M$'s need of extra
structures required to be introduced by hand by the theoretician
and not being `naturally' related to the topological manifold
({\itshape ie}, the $\cont$-continuous) one. Such structures are
the differential ({\itshape ie}, the $\smooth$-smooth) and
Lorentzian metric ({\itshape ie}, the smooth metric field
$g_{\mu\nu}$ of absolute signature $2$) ones \cite{bomb87}, and
they are implicitly postulated by the general relativist on top of
$M$'s fixed continuous topology in order to support the apparently
necessary full differential geometric ({\itshape ie}, Calculus
based!) panoply of general relativity. The $4$-dimensional,
$\smooth$-smooth Lorentzian manifold assumption for spacetime
concisely summarizes the kinematics of general relativity
\cite{sork95, mallrapt00, rapzap2, mallrapt01}.

\item (e) \underline{\bf Non-operationality:} $M$'s gravely non-operational
({\itshape ie}, non-algebraic) character as a static, pre-existent
background geometric structure---an inert stage on which fields
propagate and interact---whose existence is postulated up-front by
the theorist rather than being defined by some (algebraically
modelled) physical operations of determination or localization of
its point events. This seems to be in striking discord with the
main tenet of the philosophy of quantum theory supporting an
observer (and observation!) dependent reality \cite{rap3,
mallrapt00}. Furthermore, one would ultimately expect that {\em it
is the dynamical relations between quanta that define spacetime},
that is to say, from which spacetime, with its topological,
differential and Lorentzian metric properties, should be
effectively derived somehow \cite{rapzap2}, so that the latter
should not be regarded as an {\itshape a priori} absolute
ether-like substance \cite{einst24}---an unjustifiably necessary
passive receptacle fixed once and forever to host dynamical fields
and their interactions, but, at the same time, an entity that does
not actively participate in them\footnote{In the words of
Einstein: ``{\itshape a substance that acts, but is not acted
upon}" \cite{einst24, einst56}.}. In any case, and in view of (b)
above, we have no actual experience of a continuous infinity of
events and their differential separation cannot be recorded in the
laboratory; for evidently, realistic experiments are of finite
duration and are carried out in laboratories of finite size.
Moreover, as a matter of principle, one cannot determine the
gravitational field, hence the metric separation, between
infinitesimally separated events ({\it ie}, events whose
space-time distance is smaller than Planck's---$l_{p}\approx
10^{-35}m$-$t_{P}\approx 10^{-44}s$) without creating a black
hole. This seems to point to a fundamental cut-off of continuous
spacetime which strongly suggests that spacetime becomes reticular
or granular above a certain Planck energy ($E_{P}\approx
10^{-19}GeV$). The continuous commutatively coordinatized
geometric manifold is experimentally (or experientially!) a
non-pragmatic model of spacetime that should be replaced at a
basic level by a physically more plausible, perhaps combinatorial
and quantal ({\it ie}, noncommutative-algebraic), structure
\cite{rapzap1, rap2, mallrapt00, rapzap2, mallrapt01, rap4}.

\item (f) \underline{\bf Spatiality and globalness of topology:}
$M$'s $2$-way undirected, locally Euclidean topological structure,
will likely prove to be inadequate for modelling the irreversible
small scale connections between events, for it has been seriously
proposed that the `real' quantum theory of gravity will turn out
to be `innately' a time-asymmetric theory \cite{penr87, df88,
haag90, df911, rap4}. At the same time, the very conception of
topology as a theory of reversible, spatial (or spacelike!)
connections between points should be challenged, and justly so
because of the prominent lack of experimental evidence for
tachyons moving back and forth in spatial or spacelike directions.
In any case, the general conception of topology as the study of
the `global' features of space may seem to be problematic in a
fundamental theoresis of Physis where all significant dynamical
variables are expected to respect some kind of locality principle
({\it ie}, where all observables are in effect local variables
propagating in temporal or causal directions independently of
whether this dynamics ultimately turns out to be time-asymmetric
or not).

\end{itemize}

With these doubts about the physical soundness of the geometric
spacetime continuum in the quantum deep, we are able to discuss
next a finitary-algebraic model for (the kinematics of) discrete
Lorentzian quantum gravity that we presently possess based on
sheaf and topos-theoretic ideas, as it were, to alleviate or even
evade the aforementioned (a)-(f) `pathologies' of the classical
spacetime manifold.

\section{The Present: sheaves and their topoi in discrete Lorentzian quantum
gravity}\label{sec3}

\noindent\underline{\bf Pointlessness and Discreteness:} Finitary
substitutes for continuous spacetime topology, that is to say,
when spacetime is modelled after a topological ({\itshape ie},
$\cont$) manifold $M$, were derived in \cite{sork91} from locally
finite open covers of a bounded region $X$ of $M$ in a spirit akin
to the combinatorial \v{C}ech-Alexandrov simplicial
skeletonizations of continuous manifolds \cite{alex1, alex2,
eilsteen, dugu}. These substitutes were seen to be locally finite
$T_{0}$-posets and were interpreted as finitary approximations of
the continuous locally Euclidean topology of $M$. At the heart of
this approach to spacetime discretization lies the realistic or
`pragmatic' assumption \cite{rapzap1, rapzap2} that at a
fundamental level the singular spacetime point events should be
replaced (or smeared out) by something coarser or `larger', with
immediately obvious candidates being open sets (or generally,
`regions') about them \cite{sork91, sork95, buttish4}. Then, the
resulting poset topological spaces that substitute $M$ are, in
fact, complete distributive lattices otherwise known as locales
\cite{macmo92}. The pointlessness of these {\em finitary locales}
(finlocales) should be contrasted against the pointedness of $M$
mentioned in (a) above, while their discreteness comes to relieve
$M$'s pathology (b). Also, it should be emphasized that the
continuous $M$, with the $\cont$-manifold topology carried by its
points, can be recovered at the ideal inverse limit of infinite
refinement of an inverse system or net of these finlocales, so
that one is able to establish a connection between these discrete
topological poset substrata and the continuous manifold that they
replace on the one hand, as well as to justify their qualification
as sound approximations of $M$ on the other \cite{sork91, rap2,
mallrapt00}.

\medskip

\noindent\underline{\bf Algebra over geometry:} In \cite{rap3},
sheaves of continuous functions on the finitary locales of the
previous paragraph were studied. In the same way that the locally
finite locales were interpreted as sound approximations of the
continuous topology of (the bounded region $X$ of) the spacetime
manifold $M$, so their corresponding {\em finitary spacetime
sheaves}\footnote{We will call them `finsheaves' for short {\it
\`a la} \cite{mallrapt00}.} were viewed as reticular substitutes
of the sheaf of (algebras of) $\cont$-functions on $X$ which, in
turn, represent the {\em observables of the continuous topology of
the spacetime region $X$}. The duality of the two approaches in
\cite{sork91} and \cite{rap2} towards discretizing the spacetime
topological continuum was particularly emphasized in \cite{rap2},
namely that, while in the former scheme finitary locales
approximate the spacetime topology {\it per se}, in a more {\em
operational} ({\it ie}, {\em algebraic}) spirit that suits sheaf
theory \cite{mall1, mall2}, the latter approach discretizes (or
`finitizes') the sheaf structure of (the algebras of) our own
observations of that continuous spacetime topology. The crucial
change of emphasis in the physical semantics of the two approaches
is from discretizations of a background point set geometric realm
`out there', to ones of our very own (algebraic) operations of
perceiving that realm---which spacetime realm, for all we know,
may not physically exist independently of these operations after
all. Arguably, the finsheaf-theoretic approach comes to alleviate
the shortcoming (e) of $M$ above; while sheaves, being by
definition {\em local homeomorphisms} \cite{macmo92, mall1, mall2,
mallrapt00}, certainly address $M$'s `globalness' problem alluded
to in (f), namely, {\em local topological information is more
important than the usual global information} about, say, handles,
holes {\it etc} that (the `classical' conception of) topology is
concerned with \cite{df911, mallrapt00, rapzap2}.

\medskip

\noindent\underline{\bf Temporality and causality over spatiality
and topology:} In a dramatic change of physical interpretation of
the finitary topological posets (finlocales) involved in
\cite{sork91}, Sorkin and coworkers insisted that the partial
orders involved should not be interpreted as coarse topological or
undirected $2$-way spatial relations between geometric points, but
rather as directed $1$-way primordial causal `after' relations
between events inhabiting so-called {\em causal set}
substrata\footnote{Hereafter to be abbreviated as `causets'
\cite{mallrapt00}.} that are supposed to fundamentally underlie
the classical curved Lorentzian manifold of general relativity at
Planck scales \cite{bomb87, sork90a, sork90b, sork95, sork97,
ridesork00, sork01}. In contradistinction to the purely
topological character of the finitary locales in \cite{sork91},
causets are supposed to encode information about the microcausal
relations between events in the quantum spacetime deep. Moreover,
as Sorkin {\it et al.} stress in \cite{bomb87}, the partial order
causality relation of causets encodes almost complete information
not only about the topological ({\it ie}, $\cont$) structure of
the classical spacetime manifold, but also about its differential
({\it ie}, the $\smooth$-smooth) and conformal Lorentzian metric
structures ({\itshape ie}, the metric $g_{\mu\nu}$ of signature
$2$ modulo its determinant which represents the elementary
spacetime volume measure) that are usually externally prescribed
by the theorist on top of its continuous topology. That causality
in its order-theoretic guise is a deeper, more
physical\footnote{Especially due to lack of experimental evidence
for tachyons. Again, see section 2.} (and perhaps more pertinent
to the problem of quantum gravity) conception than topology {\it
per se} has already been amply noted in \cite{zee64, zee67, df88,
df96, sork97, rap1, mallrapt00, sork01}. The upshot of the
aforementioned `semantic reversal' is that from the causet
viewpoint, locally finite partial orders should not be viewed as
effective topological approximations of the classical spacetime
manifold, but, on the contrary, the latter should be regarded as
being of a contingent ({\itshape ie}, non-fundamental) character,
and as reflecting our own ignorance about (and related `grossness'
of our model of) the very fine structure of the world. All in all,
the manifold is the poor relative, ultimately, the approximation
of the causet, not the other way around.

To recapitulate then, partial orders as causal, not topological,
relations: this is what is `going on' between events in the
quantum deep \cite{sork01}. All this certainly presents a sound
alternative to the `additional structures' and `spatiality'
problems of the spacetime manifold mentioned above in (d) and (f),
respectively.

\medskip

\noindent\underline{\bf Finlocales and their corresponding causets
quantized:} In \cite{rapzap1}, an algebraic representation of
Sorkin's finlocales was given, namely, with every finitary poset
substitute of a continuous spacetime manifold a finite
dimensional, complex, associative and noncommutative {\em Rota
incidence algebra} \cite{rota, stanley, odon} was associated in
such a way that the topological information encoded in the former
was seen to be the same as that encoded in the latter \cite{bpz}.
Furthermore, in the new environment of the Rota algebras there is
a natural linear superposition operation between the arrows
({\itshape ie}, the partial order relations) in their
corresponding posets that is characteristically absent from
Sorkin's formulation of discrete topological spaces as posets
\cite{sork91}. In other words, and this is the main physical
interpretation of the formal mathematical structures involved in
\cite{rapzap1}, in the algebraic context one is able to form {\em
coherent quantum superpositions between the topological
connections defining these reticular topological substrata of the
classical spacetime manifold}. Moreover, this interpretation of
the incidence algebras associated with the finlocale replacements
of the classical continuum as {\em discrete quantum spacetime
topologies} enabled us to conceive of the aforementioned inverse
limit procedure by which the continuum is recovered from
finlocales in \cite{sork91} as {\em Bohr's correspondence
principle}. That is, the topological spacetime manifold arises at
the classical and experientially non-pragmatic limit of infinite
energy of resolution \cite{cole}, and concomitant `decoherence',
of an inverse system of reticular quantum topological Rota
algebraic substrata \cite{rapzap1, rapzap2}. Thus, in the Rota
finitary-algebraic context we are able to formulate a quantum sort
of spacetime topology \cite{rap4} hence evade the problematic
non-quantal nature of the continuum mentioned in (c) above.

In connection with this continuum classical limit, it should also
be mentioned that actually not only the $\cont$-topological, but
also the differential ({\itshape ie}, the $\smooth$-smooth)
structure of spacetime was anticipated in \cite{rapzap1, rapzap2}
to emerge at the classical limit from a `foam' of such discrete
quantum Rota topologies. This is so because the incidence algebras
under focus in \cite{rapzap1, rapzap2} were seen to be {\em graded
discrete differential manifolds} in the sense of Dimakis and
M\"uller-Hoissen \cite{dimu1, dimu, bdmh, dgreech}\footnote{In a
nutshell, a reticular analogue of the nilpotent K\"ahler-Cartan
differential $\dd$ (and its dual homological boundary operator
$\kodd$) can be defined on these incidence algebras. See
\cite{iasc, mallrapt01}.}. Moreover, since the differential
structure of the limit manifold represents the notion of {\em
locality} in classical spacetime physics\footnote{That is, the
local structure of classical spacetime is taken to be the point
event and the space (graded module) of differential forms
(co)tangent to it.} \cite{einst24}, these algebraic discrete
quantum topological substrata were coined `{\em alocal
structures}'---as mentioned earlier, in a sense neither local
(general relativity) nor non-local (quantum mechanics) structures
\cite{rapzap1, rapzap2}.

A couple more things should be mentioned now that we are talking
about the Rota algebraic quantization of Sorkin's finitary
locales. First, one should emphasize that the general method (and
philosophy!), originally due to Gelfand, of extracting points from
algebras as well as of assigning a fairly `natural' topology to
the latter, thus `geometrizing', as it were, algebraic structures,
was first used by Zapatrin in \cite{fas} for gathering useful
geometrical information from finite dimensional incidence algebras
and for establishing their topological equivalence to the finitary
posets of Sorkin \cite{bpz}. At the heart of this so-called
`spatialization procedure' lies the recognition that points in
these algebras are precisely the (kernels of equivalence classes
of) irreducible (finite dimensional Hilbert space) representations
of these algebras which, in turn, may be identified with the
elements of their primitive spectra ({\it ie}, the primitive
ideals in the algebras) \cite{rapzap1, rapzap2, rap4}.

The second thing that should be noted here is the categorical
duality ({\itshape ie}, a contravariant functor) between the poset
category of incidence Rota algebras associated with the finitary
locales of Sorkin, and the poset category of the latter when
viewed as simplicial complexes {\itshape \`a la}
\v{C}ech-Alexandrov\footnote{See \cite{alex1, alex2, eilsteen,
dugu} for this so-called `nerve construction of simplicial
complexes'.} \cite{rapzap1, iasc, rapzap2, mallrapt01, rap4}. In
\cite{rap4} the latter category, consisting of finitary posets or
simplicial complexes and `refinement arrows'
$\preceq$\footnote{That is, injective simplicial maps or injective
poset morphisms, or even, `continuous injections' between finitary
locales.}, was called the Alexandrov-Sorkin
category\footnote{Symbolized by $\as$.}, while the former
category, consisting of finite dimensional incidence algebras and
`coarsening arrows' $\succeq$\footnote{That is epi incidence
algebra homomorphisms.}, was coined the Rota-Zapatrin
category\footnote{Symbolized by $\rz$.}. For the time being we
note that this contravariant functor between $\as$ and $\rz$ may
be immediately recognized as defining {\em a presheaf of finite
dimensional incidence algebras over finitary locales}.

We should also mention that in \cite{rap2} an algebraic
quantization procedure of the locally finite poset structures
representing causets of Sorkin {\itshape et al.} \cite{bomb87} was
suggested based on the analogous process of quantization of
finlocales of Sorkin \cite{sork91} proposed in \cite{rapzap1}. In
little detail, with every causet its incidence Rota algebra was
associated and interpreted as a {\em quantum causal
set}\footnote{Hereafter to be referred to as `{\em qauset}'.} in
such a way that the local causal-topological information encoded
in the causet\footnote{That is, the info in the so-called
`covering relations'---the immediate causal arrows of the
underlying poset Hasse graphs.} corresponds to the one encoded in
the generating relations of the algebraic Rota topology of the
qauset\footnote{This observation will be of crucial importance in
the next paragraph where we will talk about finsheaves of qausets
as local homeomorphisms between causets and qausets
\cite{mallrapt00}.}. Another important thing to notice from
\cite{rap2} in the Rota algebraic environment that we have cast
causets, and this is the main virtue of qausets that essentially
qualifies them as the quantum analogues of causets, is that the
model allows for coherent quantum superpositions between the
causal arrows---a feature that was prominently absent from the
purely poset categorical (arrow semigroup) structures modelling
causets. Thus, we have in our hands a finitary-algebraic model for
quantum causal topology \cite{rap4}.

\medskip

\noindent\underline{\bf Curving a noncommutative topology for
qausality:} In \cite{mallrapt00}, curved finsheaves of qausets
were defined as principal finsheaves of the non-abelian incidence
Rota algebras modelling qausets having for structure group of
local symmetries a finitary version of the continuous
orthochronous Lorentz group and for base or localization space
Sorkin {\it et al.}'s causets\footnote{Technically speaking,
finsheaves of qausets over causets are local homeomorphisms
between the base causets and the qauset stalks \cite{mallrapt00}.
The aforementioned local topological equivalence between
finlocales and their incidence algebras comes in handy for
defining such finitary local homeomorphisms (finsheaves).}.
Non-trivial spin-Lorentzian ({\it ie}, $sl(2,\com)$-valued)
connections on these finsheaves were defined {\it \`a la} Mallios
\cite{mall1, mall2, mallrapt01}, and the resulting structures were
interpreted as finitary, causal and quantal substitutes of the
kinematics of Lorentzian gravity since an inverse system of these
finsheaves was seen to `converge' in the limit of infinite energy
of localization to the Lorentzian manifold---the kinematical
structure of general relativity \cite{mallrapt00}.

Then, it has been recently speculated \cite{mallrapt00, rap5,
mallrapt01, rap4} that as $\copos$---the topos of sheaves of sets
over a spacetime manifold $X$---may be viewed as a (mathematical)
universe of variable sets varying continuously over $X$
\cite{macmo92}, so a possible topos-organization of the curved
finsheaves of qausets in \cite{mallrapt00} (call it
$\qopos(\vec{P})$\footnote{The topos of sheaves of (f)initary,
(c)ausal and (q)uantal sets (qausets) over Sorkin {\it et al.}'s
causets $\vec{P}$ \cite{mallrapt00}.}) may be regarded as a
(physical) universe of dynamically variable qausets varying under
the influence of a locally finite, causal and quantal version of
Lorentzian gravity. In such a possible model it would be rather
natural to address the question opening the present paper since
the internal intuitionistic-type of logic of $\qopos(\vec{P})$
should be intimately related to the intuitionistic logic that
underlies quantum logic proper in its topos-theoretic guise
\cite{buttish1, buttish2, buttish3}. This gives us significant
hints for the deep connection between the quantum logical
structure of the world and its dynamically variable reticular
causal or chronological structure at Planck scales {\it
vis-\`{a}-vis} quantum gravity.

The discussion above brings us to the use of presheaves and their
topoi in quantum logic proper and, {\it in extenso}, to the logic
of consistent-histories.

\medskip

\noindent\underline{\bf Presheaves and their topoi in quantum
logic and consistent-histories:} In \cite{buttish1, buttish2,
buttish3}, the Koch- en-Specker theorem of quantum logic was
studied from a topos-theoretic perspective. In particular, it was
shown that quantum logic is `warped' or `curved' relative to its
Boolean sublogics \cite{rawse00}. This was achieved by showing
that certain presheaves of sets over the base poset category of
Boolean subalgebras of a quantum projection lattice $\mathcal{L}$
associated with the Hilbert space $\mathcal{H}$ (of dimensionality
greater than $2$) of a quantum system do not admit global
sections, but they do so only locally. Since these sections were
interpreted as valuations (on propositions represented by the
projectors in $\mathcal{L}(\mathcal{H})$), and since the
presheaves were organized into a topos (of so-called `varying
sets' \cite{lawv75}), the aforesaid warping phenomenon could be
read as follows: unlike classical (Boolean) logic---which is the
internal logic of the `classical' topos $\mathbf{Set}$ of constant
sets, quantum logic does not admit a global notion of truth; or
equivalently: in quantum logic truth is localized on (or
relativized with respect to) the Boolean logics embedded in it.
Furthermore, as a result of this, and as befits the internal logic
of the topos of presheaves of sets over a poset category
\cite{gold77, lambek, sel91, sel98, macmo92}, Butterfield {\it et
al.} show that quantum logic is locally intuitionistic
(`neorealist'), not Boolean (`realist').

Very similar to the treatment of quantum logic by presheaf and
topos-theoretic means above is Isham's assumption of a
topos-theoretic perspective on the logic of the
consistent-histories approach to quantum theory \cite{isham97}.
Briefly, Isham showed that the universal orthoalgebra $\latt$ of
history propositions admits non-trivial localizations or
`contextualizations' (of truth) over its classical Boolean
subalgebras. More technically speaking, it was shown that one
cannot meaningfully assign truth or semantic values to
propositions about histories globally in $\latt$, but that one can
only do so locally, that is to say, when the propositions live in
certain Boolean sublattices of $\latt$---the classical sites, or
`windows' \cite{buttish1, buttish2, buttish3}, or even `points'
\cite{mulpel1, mulpel2, rap4} within the ortholattice $\latt$.
Moreover, the simultaneous consideration of {\em all} such Boolean
subalgebras and {\em all} consistent sets of history propositions
led Isham to realize that the internal logic of the
consistent-histories theory is neither classical (Boolean) nor
quantum proper, but intuitionistic\footnote{As alluded to above in
the context of quantum logic proper, Isham in \cite{isham97} uses
the epithet `neorealist' for the quantal logic of the
consistent-histories theory in its topos-theoretic guise. Quite
resonably, we feel, one could also coin this logic `neoclassical'
\cite{rap4}---this name referring to the departure of the
Brouwerian logic of the topos of consistent-histories in
\cite{isham97} from the two-valued Boolean lattice calculus obeyed
by the states of a classical mechanical system which are modelled
after point subsets of its phase space. Arguably then, the logic
of a classical mechanical system is Boolean like that of the topos
$\mathbf{Set}$ of constant `classical' sets.}. This result befits
the fact that the relevant mathematical structure involved in
\cite{isham97}, namely, the collection of presheaves of sets
varying over the poset category of Boolean sublattices of $\latt$,
is an example of a topos \cite{lawv75, bell88, macmo92}, for it is
a general result in category theory that every topos has an
internal logic that is strongly typed and intuitionistic
\cite{gold77, lambek, sel91, sel98, macmo92}. As in the case of
quantum logic, the logic of consistent-histories is (locally)
intuitionistic (Heyting) and `warped' relative to its `local'
classical Boolean sublogics \cite{rawse00}\footnote{This departure
of quantum logic proper \cite{buttish1, buttish2, buttish3} and of
the quantal logic underlying consistent-histories \cite{isham97}
from classical Boolean logic is certainly less striking than the
famous `global' difference between quantum and Boolean logic,
namely that, while the latter is distributive, the former are
non-distributive. Thus, properly speaking, quantum logic, although
it is globally non-distributive, locally it is so; albeit,
non-Boolean, but intuitionistic (Heyting).}.

Furthermore, in \cite{rap5}, the base poset category of Boolean
sublattices of $\latt$ was endowed with a suitable Vietoris-type
of topology so that the presheaves of varying sets over the
Boolean subalgebras of $\latt$ were appropriately converted to
sheaves and, as a result, their respective topos was viewed as a
mathematical universe of sets varying {\em continuously} over
$\latt$. Moreover, the stalks of these sheaves were given further
algebraic structure---that of incidence Rota algebras---together
with the latter's physical interpretation as qausets \cite{rap2,
mallrapt00}, so that we arrived at {\em sheaves of
consistent-histories of qausets}. As a result, the topos-like
organization of these sheaves---the {\em topos of quantum causal
histories}---was anticipated to be the natural
physico-mathematical universe in which `curved quantum causality
meets warped quantum logic'\footnote{See title of the talk
delivered at QS5 (read first footnote in this paper).}, as it
were, to answer to the first question opening the present paper.

Now, the discussion about the topos of quantum causal histories
brings us to speculate briefly about the immediate future
development of our general research project `presheaves, sheaves
and their topoi in quantum gravity and quantum logic'.

\section{The Future: envisaging `quantum sheaves' and their `quantum
topoi'}\label{sec4}

Of great interest to us for the future development of our research
program, and keeping in mind the second question opening this
paper, is the following project: since the incidence algebras
modelling qausets are graded non-abelian Polynomial Identity (PI)
rings, it would in principle be possible to develop a
noncommutative sheaf or scheme type of theory \cite{harts,
eisenhar, shaf94} for such finitary non-abelian PI ring
localizations. Rigorous mathematical results, cast in a general
categorical setting, from the noncommutative algebraic geometry of
similar non-abelian schematic algebras and their localizations
\cite{voyve, voysta} are expected to deepen our physical
understanding of the dynamically variable noncommutative quantum
causal Rota topologies defined on the primitive spectra of qausets
\cite{rapzap1, mallrapt00, rapzap2, rap4}\footnote{I wish to thank
Professor Fred Van Oystaeyen (Antwerp University, Belgium) for
motivating such a study in a crucial and timely private
communication, and in two research seminars---see \cite{voystb}.}.
Ultimately, the deep connection for physics is anticipated to be
one between such a noncommutative conception of the causal
topology of spacetime and the fundamental quantum time-asymmetry
expected of the ``{\it true quantum gravity}'' \cite{penr87, df88,
haag90}. The deep connection for mathematics is, as we briefly
mentioned in section \ref{sec2}, that such a general conception of
a `noncommutative topology' is supposed to be the precursor to
Connes' `noncommutative geometry'\footnote{Freddy Van Oystaeyen in
private correspondence.} \cite{connes94}---a theory that in the
last five years or so has become of great interest to theoretical
physics, because it appears to shed more light on the persisting
problem of quantum gravity. For we emphasize again: it seems
unreasonable to have a full fledged noncommutative geometry and
lack a noncommutative topology and its corresponding sheaf theory,
especially to apply the latter to quantum gravity where even the
spacetime topology is expected to be subject to quantum dynamical
fluctuations and coherent superpositions \cite{rapzap2}.

This `noncommutative quantum causal topology' project has revived
this author's doctoral interests and work in topoi, their possible
quantization and the application of the resulting `quantum topoi'
to the problem of the quantum structure and dynamics of spacetime
\cite{rap1}. In particular, but briefly, Finkelstein \cite{df96},
as part of an ongoing effort to find a quantum replacement for the
spacetime manifold of macroscopic physics, has developed a theory
of quantum sets, which in a sense represents a quantization of
ordinary `classical' set theory. The basic idea is that spacetime
at small scales should really be viewed as a `quantum' set, not a
classical one. This is supposed to be a step on the path to a
`correct' version of quantum gravity and quantum spacetime
topology \cite{df912}. A question which may occur to a modern
logician or `toposopher' is: what is so special about the category
$\mathbf{Set}$ of classical constant sets, since there are other
logical universes just as good, and possibly better, namely
`topoi'? Perhaps it would be a better idea to try and quantize
these more general categories, since the use of $\mathbf{Set}$ may
be prey to classical chauvinism.

The usual flat ({\it ie}, in the absence of gravity) classical and
quantum field theories are conveniently formulated in
$\mathbf{Set}$, or more precisely, in $\mathbf{Sh}(X)$-the
`classical' topos of sheaves of sets varying continuously over the
classical spacetime manifold $X$ \cite{sel91, sel98}. However, as
we said in section \ref{sec2}, these theories suffer from
non-renormalizable infinities coming from singularities that
plague the smooth spacetime continuum. The manifold model, as an
inert classical pointed geometric background continuum on which
fields propagate and interact, must at least be revised in view of
the pathological nature of quantum gravity when treated as another
quantum field theory\footnote{That is, `Quantum Gravity as Quantum
General Relativity'.} \cite{mallrapt00}. Topoi and their
topological relatives, locales, which are pointless topological
spaces, are structures well-suited not to significantly commit
themselves to the pathological geometric point-like character of a
base spacetime manifold. As it has already been pointed out,
perhaps one could arrive at the `true' topos of Nature, on which a
finite quantum theory of gravity can be founded, by considering
the pointless topos of the curved finsheaves of qausets over
Sorkin's causets, or even the topos of sheaves of quantum causal
histories, instead of their classical ancestor $\mathbf{Sh}(X)$.
This quest for the `right' quantum topos of Nature is also
expected to shed more light on the following analogy that has
puzzled mathematicians for quite some time now:

\begin{equation}\nonumber
\frac{\rm locales}{\rm quantales}=\frac{\rm topoi}{\rm
?}\footnote{Jim Lambek and Steve Selesnick in private
communication. In this `proportion' the missing denominator is
supposed to be the elusive `{\em quantum topos}' structure that we
are after.}
\end{equation}

\noindent To dwell briefly on this analogy, topologically speaking
any complete distributive lattice is called a
locale\footnote{Logically speaking, a complete Heyting algebra
\cite{gold77, lambek, bell88, macmo92}.} and it corresponds to a
generalized ({\it ie}, `pointless') topological space
\cite{macmo92}. A quantale \cite{bor1, bor2, bor3, ros90, mul,
nawaz85, mulpel1, mulpel2}, the noncommutative (quantum) analogue
of a locale, may be represented by the lattice of closed two-sided
ideals of a nonabelian (von Neumann or $C^{*}$) algebra. The
primitive spectra of non-abelian qausets \cite{rapzap1, rap2,
rapzap2}, when regarded as some sort of lattices\footnote{That is,
if on top of their partial order structure, $\cap$ and $\cup$-like
operations are defined in them as in the case of the finitary
topological spaces (finlocales) mentioned in the previous
section.}, may also be viewed as some kind of quantales---albeit,
of a finitary sort \cite{rap4}, hence our regarding the topos of
finsheaves of qausets (or the sheaves of quantum causal histories)
as a strong candidate for the elusive quantum topos.

In the same line of thought, and in connection with sheaves of
qausets over consistent-histories and their possible
topos-organization mentioned at the end of the previous section,
we would like to mention another project that we are currently
working on\footnote{In collaboration with Chris Isham.}. One may
recall that in Isham's version of the quantal logic of
consistent-histories \cite{isham94, isham97} central role is
played by the tensor product `$\otimes$' structure. A sheaf $\hil$
of Hilbert spaces $\mathcal{H}$ over a classical spacetime
manifold $X$ was initially expected to be the appropriate
mathematical structure to model Isham's scenario. However, the
tensor product $\otimes$ and the `classical' definition of a sheaf
(of tensor product $\mathcal{H}$-spaces\footnote{Such sheaves may
be coined `Fock sheaves' for obvious reasons.}) do not seem to go
hand in hand for the following, at least from a physical point of
view, reason: when one considers the tensor product of two
distinct stalks in a vector sheaf like $\hil$, as when one
combines two distinct quanta in the usual quantum
theory\footnote{In the sheaf $\hil(X)$, local states of quanta are
supposed to be represented by its local sections.}, the two stalks
`collapse' to a tensor product stalk over a single spacetime point
event of the classical base spacetime manifold $X$. This
phenomenon is characteristic in both classical and quantum field
theories (in the absence of gravity) where, when we combine or
entangle systems by tensor multiplication, their spacetime
coordinates combine by identification. ``{\it This mathematical
practice expresses a certain physical practice: to learn the time,
we do not look at the system but at the sun (or nowdays) at the
laboratory clock, both prominent parts of the episystem}''
\cite{df96}, and it should be emphasized that the episystem is
always regarded as being classical\footnote{Here, the classical
base spacetime manifold $X$.} in the sense of Bohr. Thus, we
expect that the formulation of some sort of `{\em quantum sheaf}'
is required in order to be able to model non-trivially quantum
entanglement; moreover, it is quite reasonable to assume that such
a quantum notion of a sheaf will be accompanied by an appropriate
quantum notion of spacetime topology on which such sheaves are
soldered.

We conclude this paper by mentioning another potential application
of sheaf and topos theory to quantum gravity that only lately we
have envisaged and started to comprehend in full
\cite{mallrapt01}. It concerns the possible application of sheaf
and topos theory towards formulating an abstract sort of
differential geometry {\it \`a la} Mallios \cite{mall1, mall2,
mall3, mall4} on the aforementioned curved finsheaves of qausets
or their related sheaves of quantum causal histories, as it were,
to transcribe most of the differential geometric apparatus of
$\smooth$-smooth manifolds\footnote{Arguably, the mathematical
apparatus on which general relativity relies.} to a
reticular-algebraic setting that is {\it ab initio} free from the
former's pathological infinities and incurable diseases. For
instance, we have been able to perform a finitary version of the
usual $\smooth$-smooth \v{C}ech-de Rham cohomology, as well as
initiate a finsheaf-cohomological classification of the
non-trivial finitary spin-Lorentzian connections dwelling on the
curved finsheaves of qausets in \cite{mallrapt00}. In this
context, what we would also like to work on in the immediate
future is to try to relate Mallios' Abstract Differential Geometry
\cite{mall1, mall2} and its finitary applications in
\cite{mallrapt01} with the Kock-Lawvere Synthetic Differential
Geometry \cite{laven} and its promising topos-theoretic
applications to quantum gravity \cite{buttish4}. In this respect
however, the quest has just begun.

\section*{Acknowledgments}

Exchanges with Chris Isham (Imperial College) on topoi and their
potential application to quantum gravity, with Tasos Mallios
(University of Athens) on sheaves and their possible application
to quantum gravity especially via his Abstract Differential
Geometry theory \cite{mall1, mall2, mall3, mall4}, with Chris
Mulvey (University of Sussex) on quantales and noncommutative
topology, as well as a brief but timely exchange with Bob Coecke
(Oxford) on a promising dynamical conception of quantum logic
entirely by categorical means, are all greatly appreciated. The
present work was supported by the European Union in the form of a
generous Marie Curie Individual Postdoctoral Research Fellowship
held at Imperial College, London (United Kingdom).

\end{document}